\documentclass[twocolumn,aps,prd,epsf]{revtex4}
\usepackage{epsfig}
\usepackage{color}
\usepackage{amsmath}

\newcommand{\be}{\begin{equation}}
\newcommand{\bea}{\begin{eqnarray}}
\newcommand{\ee}{\end{equation}}
\newcommand{\eea}{\end{eqnarray}}

\newcommand{\mbf}{\mathbf}

\begin{document}

\title{ 
Chiral symmetry breaking in the truncated Coulomb Gauge II.
\\ Non-confining power law potentials.}
\author{
P. Bicudo}
\affiliation{CFTP, Departamento de F\'{\i}sica, Instituto Superior T\'ecnico,
Av. Rovisco Pais, 1049-001 Lisboa, Portugal}

\begin{abstract}
In this paper we study the breaking of chiral symmetry with non-confining power-like 
potentials. The region of allowed exponents is identified and, after the previous study of confining (positive exponent) potentials, we now specialize in shorter range non-confining potentials, with a negative exponent. 
These non-confining potentials are close to the Coulomb potential, and they are also relevant as corrections to the linear confinement, and as models for the quark potential at the deconfinement transition.
The mass-gap equation is constructed and solved, and the quarks mass, the chiral angle and the quark energy are calculated analytically with a exponent 
expansion in the neighbourhood of the Coulomb potential.  It is demonstrated 
that chiral  symmetry breaking occurs, but only the chiral invariant false 
vacuum and a second non-trivial vacuum exist. 
Moreover chiral symmetry breaking is led by the UV part of the potential,
with no IR enhancement of the quark mass. 
Thus the breaking of chiral symmetry driven by non-confining  potentials differs 
from the one lead by confining potentials.
\end{abstract}
\pacs{12.38.Aw, 12.39.Ki, 12.39.Pn}
\maketitle

\section{Introduction}

The problem of the spontaneous chiral symmetry breaking  ($\chi$SB) is one of
the QCD cornerstones.
 While $\chi$SB driven by a confining potential has been studied
in detail, we now explore the effect of sorter range, non-confining, 
power-law potentials in  $\chi$SB.  

Continuing the well defined mathematical problem of studying 
$\chi$SB driven by power-law potentials,
\cite{Bicudo:2003cy},
$ V(r)= \pm{K_0}^{1+\beta}r^{\beta}$,
we now specialize in negative exponents $\beta < 0$.
For diferent values of $\beta$, we find numerically
chirally noninvariant possible vacua of te theory, 
solutions to the corresponding mass-gap equation. 
We exploit the potential model for QCD
\cite{Bicudo:2003cy}, whose origins can be traced back 
to QCD in the truncated Coulomb gauge and which is proven to be successful
in studies of the low-energy phenomena in QCD \cite{pipi}. 
This class of models can be indicated as Nambu and Jona-Lasinio type models \cite{NJL} with the current-current quark interaction and the 
corresponding form factor coming from the bilocal gluonic correlator. 
A standard approximation in such type of models is to neglect the retardation and 
to approximate the gluonic correlator by an instantaneous potential of a certain form.

\begin{figure}[t!]
\resizebox{0.5\textwidth}{!}{%
\includegraphics{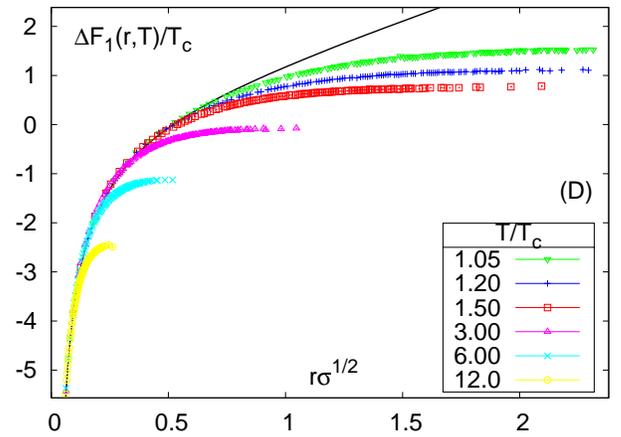}
}
\caption{\label{F1Kacz}
We show examples of non confining potentials, in particular the $T>T_c$ 
Lattice QCD data for the free energy $F_1$, thanks to 
\cite{Doring:2007uh,Hubner:2007qh,Kaczmarek:2005ui,Kaczmarek:2005gi,Kaczmarek:2005zp}Olaf Kaczmarek et al.
 The solid line represents the $T=0$ static quark-antiquark potential.}
\end{figure}

There are two different motivations to study $\chi$SB driven by
a class of non-confining potentials, the corrections to $\chi$SB 
due to non-confining potentials, and $\chi$SB at the
deconfinement transition.
Notice that the quark-antiquark static potential computed in lattice QCD has clearly two distinct  components, confinement  (linear-like ) and the shorter range Coulomb-like potential. 
In particular the shorter Coulomb-like range part of the quark-antiquark static potential may be more complicated than a pure $- \alpha \over r$ Coulomb potential. There are at least two different Coulomb potentials, the perturbative Coulomb which includes logarithmic corrections and the Luscher Coulomb due to the confining string fluctuations
\cite{Luscher:2002qv}. Moreover the matching of these two Coulomb potentials and of the long range linear potential may also be described by a non-confining potential. Thus potentials in the neighbourhood of a pure Coulomb potential, with $\beta \neq 1$
may also be phenomenologically relevant.
Moreover, at the deconfinement, say at the deconfinement phase transition 
of QCD
\cite{DeGrand:2008dh}, 
or at a large number of flavours as in walking technicolour
\cite{Belyaev:2008yj}, 
the confining potential vanishes and it is then relevant to study the impact of 
potentials, shorter range  than the confining potential, in the spontaneous 
$\chi$SB.
To illustrate that different potentials may be relevant,  in Fig. \ref{F1Kacz}
we show different finite temperature $T$ free energies computed in lattice QCD 
\cite{Doring:2007uh,Hubner:2007qh,Kaczmarek:2005ui,Kaczmarek:2005gi,Kaczmarek:2005zp}
by Kakzmarek et al. Notice that here we only address chiral symmetry breaking at zero  $T$, nevertheless the present work may also be used as a starting point for the study of $\chi$SB a finite $T$ or at finite $\mu$, where confinement is lost.

The problem of instability of the chirally invariant vacuum for power-like confining potentials, i. e. for positive exponents, has already been studied thoroughly
in the middle of 80's by the Orsay group \cite{Orsay1,Orsay2,Orsay3,Yaouanc} and such an instability was proved for the
range $0 \leq \beta<3$. 
For numerical studies, the harmonic oscillator type potential, $\beta=2$, was chosen by
these authors, as well as by the Lisbon group
\cite{Lisbon1, Lisbon2,Bicudo_scapuz}
and the Dubna 
\cite{Kalinowski}
group, 
and a set of results for the hadronic
properties were obtained in the framework of the given model. 
Adler and Davis, the Lisbon group, the Zagreb group, and the Rayleigh group also studied the linear potential many years ago
\cite{linear1,linear2,linear3,linear4}. 

Recently Bicudo and Nefediev 
\cite{Bicudo:2003cy}
solved the mass-gap equations for $0\leq \beta \leq  2$  explicitly,
and demonstrated that the chiral angle, the vacuum energy density, and the chiral condensate are smooth slow functions of the form of the confining potential, so that
the results obtained for the potential of a given form - the linear confinement being 
the most justified and phenomenologically successful choice 
\cite{linear1,linear2,linear3,linear4,Luscher:2002qv}- 
have a universal nature for any quark-quark kernels of such a type. 
Following the set of recent publications devoted to possible multiple solutions for the chirally noninvariant vacuum in QCD 
\cite{replica1} 
(see also 
\cite{replica3} 
where a similar conclusion was made in a different approach), 
Bicudo and Nefediev also addressed the question of replicas existence for various power laws 
$r^\beta$, and found that for the whole range of allowed powers, $0\leq \beta\leq 2$, 
replica solutions do exist similarly to the case of $\beta=2$ studied in detail in 
\cite{Orsay1}. 

This prompted us to extend the Coulomb potential with other negative exponents, 
and to study in detail it's contribution to $\chi$SB.  
In Section II we extend the mass gap equation for 
power-law potentials with negative exponents. 
In Section III we study analytically the mass gap equation. 
In Section IV we solve algebraicly the mass gap equation 
in the chiral limit.
In Section V we address the quark energy and the vacuum energy. 
In Section VI we conclude.

\section{The mass gap equation for the power-law potentials }

We now derive the mass gap equation for the power-law potentials. This
extends the derivation of Bicudo and Nefediev 
\cite{Bicudo:2003cy}
for positive exponents $\beta$.

The chiral model which we use for our studies is given by the Hamiltonian with the current-current
interaction parametrized by the bilocal correlator $K_{\mu\nu}^{ab}$,
\bea
\label{H}
H &=& \int d^3 \mbf x\bar{\psi}(\mbf{x},t)\left(-i
\mbox{\boldmath \( \gamma  \)}
\cdot \mbox{\boldmath \( \bigtriangledown \)}
\right)\psi(\mbf{x},t)+
\nonumber \\
&&
\frac12\int d^3 \mbf x 
d^3 \mbf y
\;J^a_\mu(\mbf{x},t)K^{ab}_{\mu\nu}(\mbf{x}-\mbf{y})J^b_\nu(\mbf{y},t),
\eea
where the quark current is $J_{\mu}^a(\mbf{x},t)=\bar{\psi}(\mbf{x},t)\gamma_\mu\frac{\lambda^a}{2}\psi(\mbf{x},t)$ 
and the gluonic correlator is approximated by a density-density potential,
\be
K^{ab}_{\mu\nu}(\mbf{x}-\mbf{y})=g_{\mu 0}g_{\nu 0}{ \delta^{ab}\over {4 \over  3}}V_0(|\mbf{x}-\mbf{y}|),
\ee
where the denominator ${4 \over3}$ normalizes the Gell-Mann matrix contribution
to the mass gap equation. We now study the class of potentials with
\be
V_0(|\mbf{r}|)= - \alpha \,{K_0}^{\beta+1}|\mbf{r}|^{\beta}\ ,
\label{potential}
\ee
where the only dimensional parameter of the model is the strength of the confining force $K_0$.

\begin{figure}[t!]
\begin{picture}(350,170)(0,0)
\put(-97,-630){\includegraphics[width=1.2\textwidth]{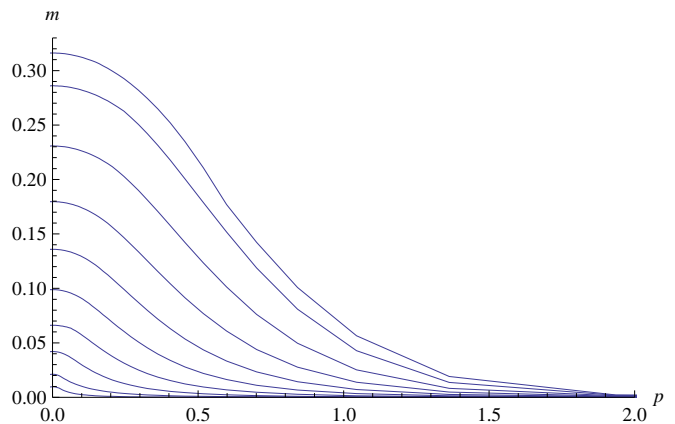}}
\end{picture}
\caption{
We show the dynamical masses $m(p)$, solutions of the mass gap equation
with zero bare mass $m_0=0$,  obtained by Bicudo and Nefediev 
\cite{Bicudo:2003cy}.
The different solutions correspond to the
 positive exponents (from left and bottom to right and top)
$\beta =$ 0.1, 0.3, 0.5, 0.7, 0.9, 1.1,
1,3, 1.5, 1.7 and 1.9.
Notice that the dynamical mass vanishes in the limit of $\beta \rightarrow 0$.
The masses are a function of the momentum $p$ and
dimensionless units of $K_0=1$ are used.
Also notice that the confining potentials enhance the masses in the IR.
}
\label{mass solutions}
\end{figure}

Previously Bicudo and Nefediev 
\cite{Bicudo:2003cy}
studied the power-law confining potentials, 
using the notation
 $ {K_0}^{\alpha+1}|\mbf{x}|^{\alpha}$, 
adequate for the study of positive exponents, in particular for
the class of confining potentials $0 \leq \alpha \leq 2$,  
including the linear and the harmonic oscillator potentials. 
The results of Bicudo and Nefediev are show in Fig. \ref{mass solutions}.

Here we specialize
in shorter range potentials, with a negative exponent. Since the Coulomb 
potential is frequently noted $- \alpha / r$, with dimensionless $\alpha$,
we now adopt the notation of eq. (\ref{potential}), where the exponent is
denoted $\beta$.  
Another difference to the previous work of  Bicudo and Nefediev 
\cite{Bicudo:2003cy}
is the sign of the potential. While the confining potentials are attractive,
it is necessary to have a negative sign for the non-confining potential 
for the potential to be attractive, and for the existence of boundstates 
in the spectrum.

The relativistic invariant Dirac-Feynman propagators
\cite{Yaouanc}, 
can be decomposed in the quark and antiquark Bethe-Goldstone 
propagators
\cite{Bicudo_scapuz},
used in the formalism of non-relativistic quark models,
\bea
{\cal S}_{Dirac}(k_0,\mbf{k})
&=& {i \over \not k -m +i \epsilon}
\nonumber \\
&=& {i \over k_0 -E(k) +i \epsilon} \
\sum_su_su^{\dagger}_s \beta
\nonumber \\
&& - {i \over -k_0 -E(k) +i \epsilon} \
\sum_sv_sv^{\dagger}_s \beta \ ,
\nonumber \\
u_s({\bf k})&=& \left[
\sqrt{ 1+S \over 2} + \sqrt{1-S \over 2} \widehat k \cdot 
\mbox{\boldmath \( \sigma \)}\gamma_5
\right]u_s(0)  \ ,
\nonumber \\
v_s({\bf k})&=& \left[
\sqrt{ 1+S \over 2} - \sqrt{1-S \over 2} \widehat k \cdot 
\mbox{\boldmath \( \sigma \)} \gamma_5
\right]v_s(0)  \ ,
\nonumber \\
&=& -i \sigma_2 \gamma_5 u_s^*({\bf k}) \ ,
\label{propagators}
\eea
where it is convenient to define,
\bea
S(k)&=&  \sin \varphi (k)=m(k ) \, D(k)
\nonumber \\
C(k)&=&   \cos \varphi (k)=k  \, D(k)
\nonumber \\
D(k) &=&  {1 \over \sqrt{k^2+{m(k)}^2}}
\eea
where  m(k) is the constituent quark mass and $\varphi$ is the chiral angle.
In the non condensed vacuum, $\varphi$ is equal to $\arctan{m_0 \over k}$.
In the physical vacuum, the constituent quark mass $m(k)$, or the
chiral angle $\varphi(k)=\arctan{m(k) \over k}$, is a variational function
which is determined by the mass gap equation. We illustrate here
examples of solutions, for the positive exponents $\beta \geq  0$
depicted in Fig. \ref{mass solutions}.

There are three equivalent methods to derive the mass gap equation
for the true and stable vacuum, where constituent quarks acquire 
the constituent mass
\cite{Bicudo:2005de}. 
One method consists in assuming a quark-antiquark $^3P_0$ 
condensed vacuum, and in minimizing the vacuum energy density. 
A second method consists in rotating the quark and antiquark 
fields with a Bogoliubov-Valatin canonical transformation 
to diagonalize the terms in the hamiltonian with two   
quark or antiquark second quantized fields. 
A third method consists in solving the Schwinger-Dyson 
equations for the propagators. Any of these methods
lead to the same mass gap equation and quark 
dispersion relation. 
Here we replace the propagator
of eq. (\ref{propagators}) in the Schwinger-Dyson equation, 
\bea
\label{2 eqs}
&&0 = u_s^\dagger(k) \left\{
k \widehat k \cdot  
\mbox{\boldmath \( \alpha \)}
 + m_0 \beta
-\int {d w' \over 2 \pi} {d^3 \mbf k' \over (2\pi)^3}
i V(k-k') \right.
\nonumber \\
&&\left. \sum_{s'} \left[ { u(k')_{s'}u^{\dagger}(k')_{s'} 
 \over w'-E(k') +i\epsilon}
-{ v(k')_{s'}v^{\dagger}(k')_{s'} 
  \over -w'-E(k')+i\epsilon} \right]
\right\} v_{s''}(k) \  \
\nonumber \\
&&E(k) = u_s^\dagger(k) \left\{k \widehat k \cdot 
\mbox{\boldmath \( \alpha \)} + m_0 \beta
-\int {d w' \over 2 \pi} {d^3 \mbf k' \over (2\pi)^3}
i V(k-k')  \right.
\nonumber \\
&&\left. \sum_{s'} \left[ { u(k')_{s'}u^{\dagger}(k')_{s'} 
 \over w'-E(k') +i\epsilon}
-{   v(k')_{s'}v^{\dagger}(k')_{s'}  
 \over -w'-E(k')+i\epsilon} \right]
\right\} u_s(k),
\eea
where, with the simple density-density harmonic interaction
\cite{Yaouanc}, the integral of the potential is a laplacian 
and the mass gap equation and the quark energy are finally,
\bea
0&=& + S(p) \,  B(p)- C(p)\, A(p) 
\label{mass gap}
\\
E(p)&=& + S(p) \,  A(p)+ C(p)\, B(p) 
\label{quark energy}
\eea
where
\bea
A(p)&=&  m_c + {1 \over 2} \int {d^3 \mathbf k \over (2 \pi)^3}  \widetilde{V}( \mathbf p - \mathbf k)S(k)
\nonumber \\
B(p)&=&  p+ {1 \over 2} \int {d^3 \mathbf k \over (2 \pi)^3}  \widetilde{V}( \mathbf p - \mathbf k)(\hat p \cdot \hat k)C(k)
\eea
Using the Bogoliubov-Valatin transformation,  
the Hamiltonian (\ref{H})  splits into the vacuum energy, the
quadratic and the quartic parts in terms of the quark creation/annihilation operators. For the vacuum energy
density one has
\bea
{\cal E}_{\rm vac}[\varphi] &=&\frac{1}{Vol}\langle 0| TH[\varphi]|0\rangle
\nonumber \\
&=&-\frac{g}{2}\int\frac{d^3 \mbf p}{(2\pi)^3}
\biggl([A(p)+ m_0] \, S(p)
\nonumber \\
&&  
\hspace{2cm}
+[B(p)+p] \, C(p)\biggr),
\label{Evac1}
\eea
where $Vol$ is the three-dimensional volume; the degeneracy factor $g$ counts the number of independent quark degrees of freedom,
\be
g=(2s+1)N_CN_f,
\ee
with $s=\frac12$ being the quark spin; the number of colours, $N_C$, is put to three, and
the number of light flavours, $N_f$, is two. Thus we find that $g=12$. 

To arrive at the mass gap equation, we compute the Fourier transform of the potential. 
To regularize the infrared (IR) part of the potential, a modified version of the potential 
( \ref{potential}) 
\cite{Orsay1}
is convenient  for $\beta \geq -1$,
\be
V_0(\mathbf r)= 
- \alpha{K_0}^{\beta+1}|\mathbf r|^{\beta}e^{-m|\mathbf r|},
\label{potential2}
\ee
where $m$ plays the role of the regulator for the infrared behaviour of the interaction, but
the limit $m\to 0$ is understood.
We get for the Fourier transform,
\bea
\widetilde{V}(\mathbf k ) 
& =& - \alpha
 \int d^3  \mathbf r  \  
{K_0}^{\beta+1}|\mathbf r| ^\beta e^{ - m r}   e^{ - i  \mathbf k \cdot  \mathbf r } 
\nonumber \\
& =& - \alpha
{ 4 \pi \over k }
\int_0^\infty dr    
{K_0}^{\beta+1}r ^{\beta+1} e^{ - m r}   \sin kr  
\nonumber \\
& =& - \alpha
{ 4 \pi \over |k| }{{K_0}^{\beta+1}\Gamma( \beta +2)\sin \left[( \beta +2)\arctan  {|k|  \over m}\right] \over (k^2+ m^2 )^{ \beta + 2  \over 2}}
\nonumber \\
& \rightarrow &
+ \alpha { 4 \pi  {K_0}^{\beta+1}\Gamma( \beta +2)\sin { \pi \beta  \over 2 }\over |k | ^{3 + \beta }}\ ,
\label{fourier}
\eea
where the Fourier transform only exists for $\beta > - 2$. For smaller exponents
the Fourier transform is UV divergent. 

For the generalized power-like potential (\ref{potential2}), 
the angular integrals necessary to compute the intermediate functions $A(p)$ and $B(p)$ are,
\bea
I_0 &=&\int_{-1}^1 d \omega  \ \widetilde{V}({k}^2 +{k'}^2 - 2 \omega k \, k') 
\nonumber \\
 &=& + \alpha    4 \pi  {K_0}^{\beta+1}\Gamma( \beta +2)\sin { \pi \beta  \over 2 }\times
\nonumber \\
 & & \ \ \left\{-{ 2 \over (1+\beta) }{ 1 \over  2 k k' }\left[{ 1 \over |k+ k' |^{1+\beta} }- {1 \over|k-k'|^{1+\beta} }\right] \right\}
\ ,
\nonumber \\
I_1 &=&\int_{-1}^1 d \omega  \ \widetilde{V}({k}^2 +{k'}^2 - 2 \omega k \, k') \ \omega 
\nonumber \\
 &=& + \alpha    4 \pi  {K_0}^{\beta+1}\Gamma( \beta +2)\sin { \pi \beta  \over 2 }\times
 \nonumber \\
 & & \ \
\Biggl\{  {2 \over (1 + \beta ) }{ 1 \over  2 k k' }\left[ {1 \over |k+ k' |^{1+\beta} }+  {1 \over |k- k' |^{1+\beta} }\right]
 \nonumber \\
 && \ \ 
 +{4 \over( -1+\beta)(1+\beta) }{ 1 \over  (2 k k')^2 }
\nonumber \\
 && \ \  \  
\left[ {1 \over |k+ k' |^{-1+\beta} }- {1 \over |k- k' |^{-1+\beta} }\right]  \Biggr\}
\ . 
\eea
and we get,
\bea
{\cal C} &=& + \alpha   { {K_0}^{\beta+1}\over 2 \pi } \Gamma( 1+\beta )\sin { \pi \beta  \over 2 }
\nonumber \\
A(p)&=&  m_0+ {\cal C} \int _{-\infty}^\infty{k}^2d  k 
\nonumber \\
&& \ \
\times
\left\{{ 1 \over  p k }\left[{ 1 \over |p- k |^{1+\beta} }\right] \right\}m(k)D(k)
\nonumber \\
B(p)&=&  p+ {\cal C}\int _{-\infty}^\infty{k}^2 d  k 
\Biggl\{  { 1 \over   p k }\left[ +  {1 \over |p- k |^{1+\beta} }\right]
 \nonumber \\
 && \ \
 +{ 1 \over   p^2 k^2 }\left[ - {1 \over |p- k |^{-1+\beta} }\right]  \Biggr\}
k D(k)
\eea
where, for the sake of convenience, we continued the integral to the negative values of $k$ assuming that $m(p)$ is an even function, as it would happen 
in 1+1 dimensions. 

Consequently, the mass-gap equation (\ref{mass gap}) takes the form  
of a non-linear integral equation for the constituent mass $m(p)$
\bea
m(p)&=&  m_0  +  \int _{-\infty}^\infty d  k   \ {\cal C}{ 1 \over   p^3 }
D(k)
\nonumber \\
&& \ \ \
\times
\Biggl\{  +\left[   {p k \over |p- k |^{1+\beta} }\right]  \left[ \, p \, m(k)- \, k \, m(p)\right]
 \nonumber \\
 && \ \ \
 -{1 \over( -1+\beta) }\left[ {1 \over |p- k |^{-1+\beta} }\right] \, k \, m(p) \Biggr\} 
\label{masseq}
\eea
and this is the main object of our studies.

\section{Analytical properties of the mass gap equation }

We now analyse dimentionally the mass gap equation, and study possible
infrared (IR) and ultraviolet (UV) divergences.

In Section II the possible exponents $\beta$ are already limited
to $\beta > - 2$, since in eq. (\ref{fourier}) the Fourier 
transform of the potential does no exist for smaller exponents. 
Now we address in a dimensional analysis the stability of a possible 
non-trivial vacuum. Let us assume that a solution $m(k)$ exists, 
minimizing the vacuum energy. Then we arbitrarily rescale the solution,
\be
m(k) \rightarrow m(\kappa \, k)
\ee
and, if the vacuum energy does not have an absolute minimum for
$\kappa =1$, the vacuum is unstable and thus our assumption was wrong.
We may now  simply evaluate the vacuum energy as a function of the 
dimensionless factor $\kappa$. We get, from eq. (\ref{Evac1}),
\be
{\cal E}_{vac}(\kappa)= c_1 \kappa^4 + c_2 m_0 \kappa^3 + c_3{K_0}^{\beta+1}\kappa^{-\beta+3}
\label{energyscaling}
\ee
the vacuum energy density, with a dimension of the fourth power of momentum.
In eq. (\ref{energyscaling}), the $c_i$ are constants, equal to the different  integrals in the vacuum energy density (\ref{Evac1}).
Thus, in the chiral limit which is the one mattering here, the kinetic
energy density scales like $\kappa^4$. We can show that $c_1$ is positive, 
and this prevents the vacuum to be UV unstable, providing the potential 
term has a smaller scaling power than the kinetic energy density. Thus
for $\beta > -1$  there may be a solution. The nicer case is the one of
the linear potential where the  ${\cal E}_{vac}(\kappa)$ has 
a perfect Mexican hat shape. For  $\beta < -1$, the potential always
wins the kinetic term, moreover for an attractive potential we can show that
the constant $c_3$ is negative, and thus the vacuum is unstable.
For the Coulomb case, $\beta = -1$ both terms scale equally, and there
is either a trivial solution $m(k)=0$ if the $c_3 < c_1$ 
or the vacuum is unstable if $c_1>c_3$.
Thus we show that there may be a stable and non-trivial 
$m(k) >0$ solution to the mass gap equation only for $\beta > - 1$.  
Since the present
paper is specialized in negative exponents, we are interested in
solving the mass gap equation for  $-1 < \beta  < 0$. 

In the present case of a negative exponent $\beta$, we now
show that the gap equation (\ref{mass gap})is IR finite. 
Notice that in the
mas gap equation any possible IR divergence may only occur in the two 
denominators with a power of $|p-k|$.
The second denominator $ 1 \over |p-k|^{-1+\beta}$ is clearly IR finite
since the exponent $1-\beta >1$. 
The first denominator $ 1 \over |p-k|^{1+\beta}$ tends to an IR divergence 
when $\beta \rightarrow 0$, however this divergence is cancelled by the 
numerator  $p \, m(k) -k \, m(p)$. Thus the mass gap equation 
(\ref{mass gap})  for negative exponents is quite different from the
mass gap with positive exponents, where an exact cancellation of the
IR divergences of these two different terms would occur, but nevertheless
would be technically harder to implement in the mass gap equation.

In the present case of IR finiteness, we now focus in the UV sector
of the equation. 
In what concerns the UV limit, each separate term in the integrand of the
mass gap equation (\ref{mass gap}) may be UV divergent in the limit of 
$\beta \rightarrow -1$, when the integrand momentum $k$ tends to $\pm \infty$. 
To study whether the UV divergences cancel, it is convenient to perform momenta expansions in the integrand. 
There are two different expansions of interest, one where
the momentum in the integrals is much larger than the external momentum, corresponding to the the limit of $|k| >> |p|$,  where the mass is not limited. A second possible limit is the one of $k, p >> m$ when we are interested in
large external momenta, and we assume that the mass is limited. We 
start by expanding the integrand  in the limit when $ {k \over p }\to 0$.

In order to utilize as much as possible the cancellations of UV divergences 
of the different terms in the mass gap equation, we not only sum all
the terms but also return to a momentum integral from $0$ to $\infty$. Then,
if $I(k)$ is the integrand of the mass gap eq. (\ref{masseq}), for the
expansion in  $k \over p $ of the mass independent terms we get,
\bea
&& I(k)+I(-k)=  { \alpha  \over 2 \pi}\Gamma(2+ \beta )\sin { \pi \beta  \over 2 }{{K_0}^{1+\beta} D(k)\over  |k|^{1+\beta} }\times
\nonumber \\
&& \hspace{.5cm} \Biggl\{ 
{2 \over 3}\left[ (3+ \beta ) m(p)-  3 m(k)\right]
 +  {1 \over 15} (3 + \beta )(2 + \beta)  \times
\nonumber \\
&& \hspace{.5cm} 
\left[  (5+ \beta)m(p) - 5 m(k) \right] {p^2 \over k^2}
+o({p^4 \over k^4})
 \Biggr\}
\eea
thus, at leading order in $k \over p$, the mass gap equation
(\ref{masseq}) can be rewritten as,
\bea
\label{UVmasseq}
{\cal C}' &=& { \alpha  {K_0}^{1+\beta} \over  2 \pi }
\Gamma(2+ \beta )\sin { \pi \beta  \over 2 }{2 \over 3}
\\ \nonumber
m_0 &=& m(p)- \int_0^\infty dk \,  {\cal C}'  \,
{ D(k)\over  |k|^{1+\beta} } \, \left[ (3+ \beta ) m(p)-  3 m(k)\right]
 \  .
\eea
The eq. (\ref{UVmasseq}) is UV divergent for $ \beta  \leq -1$
but it is indeed UV finite for  $-1 < \beta  < 0$, and thus we can proceed
in our search of it's solution. 

\section{Algebraic solution of the mass gap equation in the chiral limit}

We now utilize algebraic methods to solve the mass gap equation. 
We first set the mass gap equation in a form close to an eigenvalue 
equation and show that the eigenvalues of this equation are real.
The solutions of the mass gap equation are the roots of the linearized
eigenvalue equation, and this provides a fast convergence to the solution. 
We provide the solution in the limit where the UV contribution leads the 
mass gap equation.

We consider the chiral limit of $m_0\rightarrow 0$.
We first address the mass gap equation in three momentum dimensions (3d).
The 3d mass gap  equation  (\ref{mass gap}) van be rewritten as
\bea
0&= &\left[ 1  \over D(p)\right]  D(p)\, m(p) 
\nonumber \\
&&
+  \left[
{1 \over 2} \int {d^3 \mathbf k \over (2 \pi)^3}  \widetilde{V}( \mathbf p - \mathbf k)(\hat p \cdot \hat k)\, {k\, D(k)\over p \, D(p)}
\right]
 D(p) \, m(p)
\nonumber \\
&&
\left[ 
-{1 \over 2} \int {d^3 \mathbf k \over (2 \pi)^3}  \widetilde{V}( \mathbf p - \mathbf k) 
\right]
 D(k)\, m(k)
\label{symmetric3d}
\eea
and this is similar to a 3d - like algebraic quasi-linear equation with a symmetric matrix for a vector $D(k) m(k) $, since the integrand only depends on the
distance $\mbf p -  \mbf k$. 

To solve the mass gap equation,
we start by fixing the denominator functions $D(k)$ 
inside the square brackets $[\ ]$ of eq. (\ref{symmetric3d}),
with our best initial guess. Then we apply the eigenvalue method 
to the resulting matrix equation.  Clearly, the eigenvalues are real 
since the matrix is real and symmetric. A solution 
of the mass gap exists when the matrix has a root. In eq. (\ref{symmetric3d}) we have three matrices. The first one is the identity and has positive
eigenvalues growing with the mass function $m(p)$. 
The other two matrices have eigenvalues with little dependence on  
the mass function $m(k)$. 
If, in the limit of vanishing $m(k)$, there are one or more negative eigenvalue, then when we increase the mass $m(k)$, the first matrix increases and eventually it is able to cancel the negative eigenvalues of the second plus third matrices. In that case we find the 
desired roots, and we solve the mass gap equation. The number of solutions 
is equal to the number of negative eigenvalues of the matrix computed in
the massless limit.

To actually solve the mass gap equation it more convenient to solve
the radial momentum version  (1d) of the mass gap equation.
For the 1d mass gap we can start from eq.  (\ref{masseq}), 
and rewrite it is a symmetric form,
\bea
0 &=&  \left[1 \over D(p)\right]  
D(p)\, p \, m(p)
+ \int _{-\infty}^\infty d  k {{\cal C} \over  p^3 }
k {D(k)\over D(p)}
\,
\Biggl[ +  {p k \over |p- k |^{1+\beta} } 
 \nonumber \\
 && \ \ \
 +{1 \over( -1+\beta) }{1 \over |p- k |^{-1+\beta} }\Biggr] 
D(p)\, p \, m(p)
\nonumber \\
 && +  \Biggl[  \int _{-\infty}^\infty d  k {\cal C}
   { -1 \over |p- k |^{1+\beta} }  \Biggr]  \, D(k)\, k \, m(k)
\eea
and again particular we have two terms, one fully diagonal, and another explicitely
symmetric since it only depends on the diagonal distance
$|p-k|$. Thus the matrix is hermitean, and the eigenvalue equation
now applies to a vector  $D(k) \, k \, m(k)$.  Again, the number of 
solutions is equal to the number of negative eigenvalues of the matrix 
computed in the massless limit of $ D(k) \to {1 \over k}$.

\begin{figure*}[t!]
\begin{picture}(500,180)(0,0)
\put(-100,-630){\includegraphics[width=1.2\textwidth]{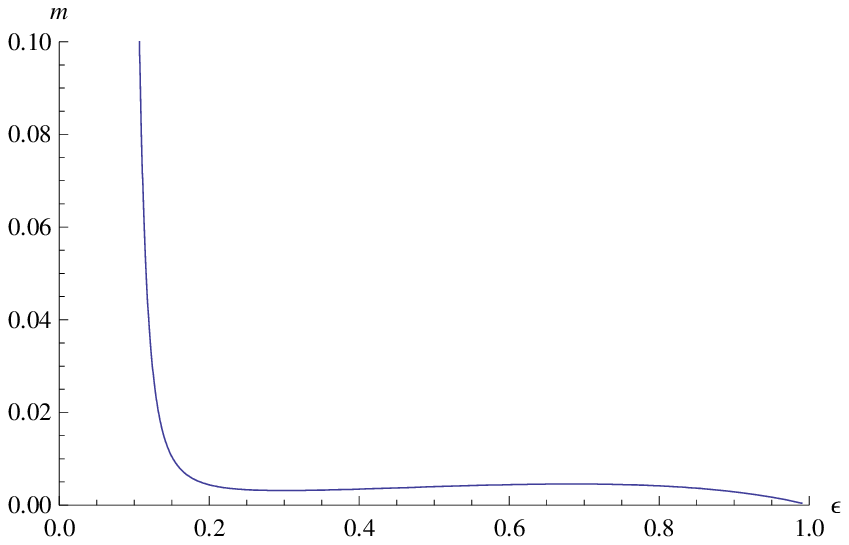}}
\put(165,-630){\includegraphics[width=1.2\textwidth]{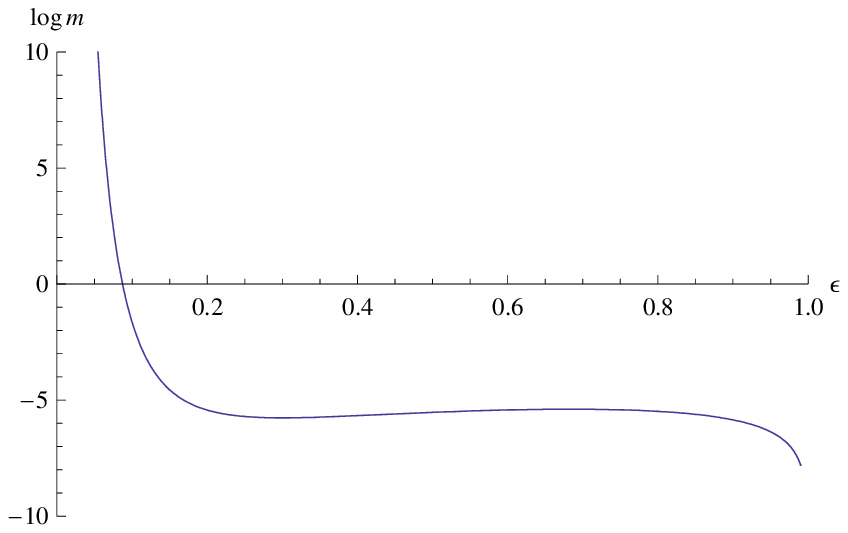}}
\end{picture}
\caption{We show, (left)  the quark mass $m$, and (right) $log(m)$, 
solution of the mass gap equation in the chiral limit, plotted as a 
function of the  exponent $\epsilon$, in dimensionless units of $K_0=1$. 
We 
consider here $\alpha = \pi / 12$ as in the Luscher term 
\cite{Luscher:2002qv}
computed in static Lattice 
QCD potentials.} 
\label{Mass of epsilon}
\end{figure*}

We now solve the mass gap equation in the limit where it is lead
by the UV contribution.
From the UV lead mass gap eq. (\ref{UVmasseq})we get the matrix equation,
\bea
0 &=& \left[ 1- \int_0^\infty dk \,  {\cal C}'  \,
{ D(k)\over  |k|^{1+\beta} } \, (3+ \beta )
\right ] \, m(p) 
\nonumber \\
&& +  \left[  \int_0^\infty dk \,  {\cal C}'  \,
{ D(k)\over  |k|^{1+\beta} } \, 3
\right ] \, m(k) \  .
\label{mgeUV}
\eea
Let us consider that we discretize the momenta, for instance in a lattice of $N$ points, where the correct solution is found in the limit $N \to \infty$. 
Then the  mass gap equation (\ref{mgeUV})
is a $N \times N$ matrix equation. In this case the corresponding matrix is not symmetric, but actually we can find the eigenvalues exactly and show that
they are real,  since one of the matrices is a constant and the other is a projector. 

In particular,
 the eigenvalue equation applies to the vectors  $m(k)$.
The first term in eq. (\ref{mgeUV}), proportional  to $m(p)$,
is constant, and is thus proportional to the identity matrix.
The term integrating in $m(k)$ does not depend on $p$ and thus it is a projector on a constant vector $m_1(p)= cst$.

This implies that one eigenvector, say $\lambda_1(p)$,  of the $N \times N$ matrix 
is constant,
\bea
\lambda_1 &= & \left\{ 1- \int_0^\infty dk \,  {\cal C}'  \,
{ D(k)\over  |k|^{1+\beta} } \, (3+ \beta )
\right \}
\nonumber \\
&& +  \left\{ \int_0^\infty dk \,  {\cal C}'  \,
{ D(k)\over  |k|^{1+\beta} } \, 3
\right \}
\nonumber \\
&=& \left\{ 1- \int_0^\infty dk \,  {\cal C}'  \,
{ D(k)\over  |k|^{1+\beta} } \,  \beta
\right \}\ .
\eea

And, since in the projector all the lines of the discretizing matrix 
are identical, the matrix has $N-1$ linear dependences,
  and this implies that all the other $N-1$ eigenvectors  
$\lambda_i(p)$, for $i >1$, 
are cancelled by the projector, and thus all their eigenvalues are identical, 
simply given by the matrix proportional to the identity.
The other eigenvalues $\lambda_i (p) \ , \ \ i>1$ cancelled by the projector  are,
\bea
 \lambda_i  &=& 
\left\{ 1- \int_0^\infty dk \,  {\cal C}'  \,
{ D(k)\over  |k|^{1+\beta} } \, (3+ \beta )
\right \}\ .
\eea

Now,  notice that ${\cal C}' <0$ if we consider an attractive, i.e. negative
shorter range potential with $- \alpha  <0$, 
and with a negative $\beta \simeq -1 $, and leading us to,
\bea
\lambda_1 -1 &= &  - \int_0^\infty dk \,  {\cal C}'  \,
{ D(k)\over  |k|^{1+\beta} } \,  \beta 
< 1\ ,
\nonumber  \\
\lambda_i -1&= &   - \int_0^\infty dk \,  {\cal C}'  \,
{ D(k)\over  |k|^{1+\beta} } \,  (3 + \beta )
> 1 \ .
\eea
Thus we can have one, and only one root. The other eigenvalues 
are positive (and quite large if the integral is nearly UV divergent). 
Thus this differs from the confining potentials where a whole tower of 
replicas was found
by Bicudo and Nefediev 
\cite{Bicudo:2003cy}.

We now concentrate on the root to determine the 
constant eigenvector $m$ that solves the mass gap equation
(\ref{UVmasseq}).
It is convenient to rename the exponent to
 $\beta = -1 + \epsilon$, where we  are interested in the exponent
range of $0<\epsilon <1$. The mass gap equation is then, 
\bea
0 &=& \lambda_1 \ ,
\\ \nonumber 
&=& 1 - 
\int_0^\infty dk \,  {\cal C}'  \,
{ 1 \over  \sqrt{ k^2 +m^2}|k|^{\epsilon} }
\\ \nonumber 
&=&1- {\cal C}' (-1+ \epsilon){m ^{-\epsilon}\Gamma \left( 1-\epsilon \over 2 \right) \Gamma \left( \epsilon \over 2 \right) \over 
2 \sqrt \pi  } \ ,
\eea
and the solution is,
\bea
m &= & \left[ 
\left| {\cal C}'\right|
(1- \epsilon){\Gamma \left( 1-\epsilon \over 2 \right) \Gamma \left( \epsilon \over 2 \right) \over 
2 \sqrt \pi  }\right]^{1 / \epsilon}
\nonumber \\
&= & K_0 
\left[ \alpha
{\Gamma(1+ \epsilon  )\over  2 \pi }
\sin { \pi ( 1- \epsilon )  \over 2 }
{1- \epsilon \over {3 \over 2} }
{\Gamma \left( 1-\epsilon \over 2 \right) \Gamma \left( \epsilon \over 2 \right) \over 
2 \sqrt \pi  }\right]^{1 / \epsilon}
\nonumber \\
& = & K_0 
\left[  
{\alpha \over 3 \pi \epsilon }
-  \alpha {  2 + 3 \gamma  + \psi \left(1 \over 2 \right)\over 6 \pi} 
+o(\epsilon)
\right]^{1 / \epsilon}
\nonumber \\
& = & \left| K_0 \right|
\left(\alpha \over 3 \pi \epsilon  \right)^{1 / \epsilon}
e^{\left[ { - \alpha{  2 + 3 \gamma  + \psi \left(1 \over 2 \right)\over 6 \pi} 
+o(\epsilon)}  \right]}
\label{mass divergent}
\eea
and this diverges very fast in the limit of vanishing $\epsilon$, but for 
finite $\epsilon$ it occurs that the negative $\beta$ exponents indeed
produce chiral symmetry breaking. 

The solution  of  eq. (\ref{mass divergent}) for the dynamical quark mass is plotted in 
Fig. \ref{Mass of epsilon} as a function of $\epsilon$.
We find  that a small but finite mass $m \simeq 0.05 K_0$, 
almost independent of $\epsilon$,
in the range $0.15 < \epsilon < 0.9$. For 
$\epsilon \simeq 1$, corresponding to $\beta \simeq 0$ the mass vanishes.
The mass explodes for $\epsilon \simeq 0$, close to the coulomb potential.

\section{The quark energy, in the chiral limit and with a bare mass }

To compute the quark energy, 
it is first convenient to write the mass gap equation (\ref{mass gap}) as, 
\bea
A(p) \over m(p) & = & {B(p) \over p} \ ,
\eea
and then we get for the quark energy (\ref{quark energy}),
\bea
E(p)&=& D(p) \left[ m(p) \, A(p) + p  \, B(p)\right]
\nonumber \\
&=& \sqrt{m(p)^2 + p^2}{ \, A(p) \over m(p) } 
\nonumber \\
&=& \sqrt {m ^2 + p^2}{3 \over 1 - \epsilon}
\eea
where $m$ is computed in eq. (\ref{mass divergent}).

Since $m$ is UV divergent when $\epsilon \rightarrow 0$,
In order to have a limited quark mass when $\epsilon$  is small, one 
needs to include in the mass gap equation, as a counter term, a bare 
quark mass $m_0$.  
In this  case we may use a different limit, where $m <<, p, k$, still
considering for simplicity a constant $m(p)=m$, and
we get for the $A(p)$ and $B(p)$  integrals,
\bea
{A(p)\over m}&=&  {m_0 \over m}+{\cal C} \int _{-\infty}^\infty{k}^2d  k 
\left\{{ 2 \over  2 p k }\left[{ 1 \over |p- k |^{1+\beta} }\right] \right\}{1 \over | k |}
\nonumber \\
&=&  {m_0 \over m}+{{\cal C} \over p ^ \epsilon }{2 \over 1 - \epsilon} \ ,\nonumber \\
&=&  { m_0 \over m }
+  \alpha
\left[
{ 1 \over \pi \epsilon} 
+{ 1 - \gamma - \log\left(p \over K_0 \right)\over \pi} 
+ o(\epsilon)
\right]
\nonumber \\
{B(p)\over p} &=&  1+  {\cal C} \int _{-\infty}^\infty {k}^2d  k 
\Biggl\{  { 2 \over  2 p k }\left[ +  {1 \over |p- k |^{1+\beta} }\right]
 \nonumber \\
 && \ \ \
 +{4 \over( -1+\beta) }{ 1 \over  (2 p k)^2 }\left[ - {1 \over |p- k |^{-1+\beta} }\right]  \Biggr\}
k {1 \over | k |}{1 \over p}
\nonumber \\
&=&  1 + {{\cal C} \over p ^\epsilon }{4 \over (1 - \epsilon)(3- \epsilon)}
\\ \nonumber 
&=&
1 
+ 
{2 \alpha \over 3 } \left[
{ 1 \over \pi \epsilon} 
+{ 4 - 3 \gamma - 3 \log\left(p \over K_0 \right)\over  3 \pi} + o(\epsilon)
\right]
\ .
\eea
Thus, assuming an approximately constant $m$  we get for the bare
mass $m_0$ which acts here as renormalization counter term,
\be
m_0= \left[
- {\alpha \over 3  \, \pi \, \epsilon }
+0(\epsilon^0)\right] m
\ee
and thus, if we specialize in the minimal subtraction scheme,
we get a finite mass. Nevertheless  the quark energy remains 
UV divergent in the limit of $\epsilon \to 0$,
\be
E(p)= \sqrt{ P^2+ m^2}
\left[{ 2 \alpha \over 3 \pi \epsilon}+ o (\epsilon^1)\right]
\ee

A further renormalization of the quark energy may be necessary, but
only upon studying the Bethe Salpeter equation for mesons, which
goes beyond the scope of the present paper. Different methods
to renormalize the Coulomb potential in the Coulomb gauge have been 
applied by the Zagreb group
\cite{linear3}
and by Szczepaniak and Swanson
\cite{Szczepaniak:1996gb}
utilizing the Glazek and Wilson method
\cite{Glazek:1993rc}.

\section{Conclusion }

We study $\chi$SB driven by power-law potentials
with a negative exponent $\beta < 0$. These potentials are non-confining.
We extend a previous study performed for confining potentials with a
positive exponent
\cite{Bicudo:2003cy}.
In that study, chiral symmetry already vanishes when $\beta \to 0$,
nevertheless, since the Coulomb potential is negative, it is natural  to
reverse the sign of the potential, and then $\chi$SB may occur for
$\beta < 0$.

We work in momentum space, and the existence of a Fourier transform 
is limited to $ \beta > -2$. Then, with a dimensional analysis, we show that
we may only have a stable $\chi$ S B vacuum for $ \beta \geq -1$. Thus we
study in detail the non-confining potentials with  $-1 < \beta < 0$, i. e.  in the exponent range limited by the Coulomb potential and the logarithmic potential. 

We find that $\chi$SB also occurs for the studied negative exponent
power-law potentials generating dynamically a finite quark mass $m$.

Moreover we find qualitative differences to the mass generated with
confining potentials. First, we find one and only one non-trivial solution of the
mass gap equation, whereas for confining potentials an infinite tower
\cite{Bicudo:2003cy}
of false, excited vacua, sometimes called replicas, is found. 
Also, the solution found has an approximately constant mass, 
as in Fig. \ref{Mass of epsilon}, i. e we find no IR enhancement of the quark mass, 
whereas the confining potentials studied previoulsy 
\cite{Bicudo:2003cy}
produce a significant IR enhancement of the quark mass, as in Fig. 
\ref{mass solutions}.

We also find an UV divergence of the dynamical mass $m$, in
the Coulomb potential limit of the exponent $\beta \to -1$. 
This is consistent with the well known necessity to apply a renormalization 
program when the Coulomb potential is used. 

This work may be a starting point for the study of $\chi$SB at the
deconfinement transition, say with finite $T$, finite $\mu$ or large $N_F$.

\acknowledgments
The author is grateful to Alexei Nefediev for valuable suggestions that motivated this work, 
and to Olaf Kaczmarek for sharing it's Lattice QCD static quark potentials and energies.  
This work was partly funded by the FCT grants,
PDCT/FP/63923/2005 and POCI/FP/81933/2007.


%

\begin{thebibliography}{99}

\bibitem{Bicudo:2003cy}
  P.~J.~A.~Bicudo and A.~V.~Nefediev,
  Phys.\ Rev.\  D {\bf 68}, 065021 (2003)
  [arXiv:hep-ph/0307302].



\bibitem{pipi} P. Bicudo, S. Cotanch, F. Llanes-Estrada, P. Maris, E. Ribeiro,
and A. Szczepaniak, Phys. Rev. D {\bf 65}, 076008 (2002).

\bibitem{NJL} Y. Nambu, G. Jona-Lasinio, Phys. Rev. {\bf 122}, 345
(1961).



\bibitem{Luscher:2002qv}
  M.~Luscher and P.~Weisz,
  JHEP {\bf 0207}, 049 (2002)
  [arXiv:hep-lat/0207003].


\bibitem{DeGrand:2008dh}
  T.~DeGrand, Y.~Shamir and B.~Svetitsky,
  arXiv:0809.2953 [hep-lat].

  
\bibitem{Belyaev:2008yj}
  A.~Belyaev, R.~Foadi, M.~T.~Frandsen, M.~Jarvinen, F.~Sannino and A.~Pukhov,
  arXiv:0809.0793 [hep-ph].


\bibitem{Doring:2007uh}
  M.~Doring, K.~Hubner, O.~Kaczmarek and F.~Karsch,
  Phys.\ Rev.\  D {\bf 75}, 054504 (2007)
  [arXiv:hep-lat/0702009].

\bibitem{Hubner:2007qh}
  K.~Hubner, F.~Karsch, O.~Kaczmarek and O.~Vogt,
  arXiv:0710.5147 [hep-lat].
  
\bibitem{Kaczmarek:2005ui}
  O.~Kaczmarek and F.~Zantow,
  Phys.\ Rev.\  D {\bf 71}, 114510 (2005)
  [arXiv:hep-lat/0503017].
  
\bibitem{Kaczmarek:2005gi}
  O.~Kaczmarek and F.~Zantow,
  arXiv:hep-lat/0506019.

\bibitem{Kaczmarek:2005zp}
  O.~Kaczmarek and F.~Zantow,
  PoS {\bf LAT2005}, 192 (2006)
  [arXiv:hep-lat/0510094].
  

  

\bibitem{Orsay1} A. Le Yaouanc, L. Oliver, O. Pene, J. C. Raynal, Phys. Lett.
{\bf 134B}, 249 (1984).

\bibitem{Orsay2} A. Amer, A. Le Yaouanc, L. Oliver, O. Pene and J.-C. Raynal, Phys. Rev. Lett.
{\bf 50}, 87 (1983).

\bibitem{Orsay3} A. Le Yaouanc, L. Oliver, O. Pene and J.-C. Raynal,
Phys. Rev. D {\bf 29}, 1233 (1984); 

\bibitem{Yaouanc}
A.~Le Yaouanc, L.~Oliver, S.~Ono, O.~P\`ene  and J.~C.~Raynal,
  Phys.\ Rev.\ D {\bf 31}, 137 (1985).

\bibitem{Lisbon1}  
 P. Bicudo, J. E. Ribeiro, Phys. Rev. D {\bf 42}, 1611 (1990);
{\it ibid.}, 1625 (1990); {\it ibid.}, 1635 (1990).

\bibitem{Lisbon2}
P. Bicudo, Phys. Rev. Lett.
{\bf 72}, 1600 (1994); 

\bibitem{Bicudo_scapuz}
  P.~Bicudo,
  Phys.\ Rev.\ C {\bf 60}, 035209 (1999).


\bibitem{Kalinowski}
Y. L. Kalinovsky, L. Kaschluhn and V. N. Pervushin, Phys. Lett. B {\bf 231}, 288 (1989).


\bibitem{linear1} S. L. Adler, A. C. Davis, Nucl. Phys. B {\bf 244}, 469 (1984),

\bibitem{linear2}
P. Bicudo, J. E. Ribeiro and J. Rodrigues, Phys. Rev. C {\bf 52}, 2144 (1995).

\bibitem{linear3}
R. Horvat, D. Kekez, D. Palle and D. Klabucar, Z. Phys. C {\bf 68}, 303 (1995).

\bibitem{linear4}
F. J. Llanes-Estrada, S. R. Cotanch, Phys. Rev. Lett.  {\bf 84}, 1102 (2000).


\bibitem{replica1} P.  Bicudo, A. N. Nefediev, and J. E. F. T. Ribeiro, Phys.
Rev. D {\bf 65}, 085026 (2002).

\bibitem{replica3}  A. A. Osipov, B. Hiller, Phys. Lett. {\bf 539B}, 76 (2002).





\bibitem{Bicudo:2005de}
  P.~Bicudo,
  Phys.\ Rev.\  D {\bf 74}, 036008 (2006)
  [arXiv:hep-ph/0512041].



\bibitem{Szczepaniak:1996gb}
  A.~P.~Szczepaniak and E.~S.~Swanson,
  Phys.\ Rev.\  D {\bf 55}, 1578 (1997)
  [arXiv:hep-ph/9609525].

\bibitem{Glazek:1993rc}
  S.~D.~Glazek and K.~G.~Wilson,
  Phys.\ Rev.\  D {\bf 48}, 5863 (1993).


\end{thebibliography}
\end{document}